\documentstyle [preprint,prl,aps,epsf] {revtex}

\begin{document}
\renewcommand{\[}{\begin{equation}}
\renewcommand{\]}{\end{equation}}

\draft
\title{Evidence for d-wave superconductivity in the repulsive Hubbard-model}
\author{W. Fettes and I. Morgenstern}
\address{Fakult\"at Physik, Unversit\"at Regensburg, D-93040 Regensburg\\
e-mail: werner.fettes@physik.uni-regensburg.de}
\date{10. December 1997}

\maketitle

\begin{abstract}
We perform numerical simulations of the Hubbard model
using the projector Quantum Monte Carlo method.
A novel approach for finite size scaling is discussed.
We obtain evidence in favor of d--wave superconductivity in the
repulsive Hubbard model. For $U=4$, $T_c$ is roughly estimated as
$T_c\approx 30$\,K.
\end{abstract}


\vglue1cm

After the discovery of high--temperature superconductivity (HTSC) 
the two-dimensional Hubbard model \cite{HUB63}, \cite{AND87}  has been
proposed as a model for a theoretical explanation of the phenomena.
Indeed it has been shown, that the Hubbard model exhibits the 
normal conducting and magnetic properties of HTSC \cite{MON92/2}.

It is now widely accepted, that
HTSC show d-wave symmetry of the superconducting order parameter 
\cite{WOL93}, \cite{TSU94}. 
According to our simulations \cite{MOR94}, \cite{HUS94} and recent work
\cite{ZHA97} the repulsive Hubbard model also favors d-wave symmetry.
But the question of (d--wave) superconductivity in the Hubbard 
model has been discussed controversially \cite{MOR94}, \cite{HUS94}, \cite{ZHA97}.

We proposed the tt'--Hubbard model as the suited model for
numerical simulations \cite{MOR94}, \cite{HUS94}. It exhibits a Van 
Hove singularity away from half filling \cite{HUS94}.
Furthermore with the tuning of the next nearest neighbor hopping $t'$
we are in the position to circumvent the majority of numerical 
difficulties in the simulation.
The tt'--Hubbard model is described by the Hamiltonian
\[
\label{eqhubbard}
{\cal H}^{Hub} = 
-t \sum_{<i,j>}\sum_{\sigma} c_{i,\sigma}^\dagger c^{}_{j,\sigma}
-t' \sum_{\ll i,j\gg } \sum_{\sigma} c_{i,\sigma}^\dagger c^{}_{j,\sigma} 
+
U \sum_i n_{i,\uparrow} n_{i,\downarrow} 
\quad .
\]
Here 
$c^\dagger_{i,\sigma}$ creates an electron with spin $\sigma$ on site $i$,
$n_{i,\sigma}$ is the corresponding number operator
and $U$ is the on-site Coulomb interaction. The sum $<i,j>$ ($\ll i,j\gg $) 
runs over the pairs of (next) nearest neighbors. 

Our simulations are performed with the projector quantum Monte Carlo 
method (PQMC) \cite{KOO82}, \cite{SUG86}, \cite{SOR89}.
In which the ground state 
\[
|\Psi_0 \rangle  = 
\frac{1}{{\cal N}} {\rm e}^{-\theta {\cal H}} | \Psi_T\rangle 
\]
of the Hamiltonian ${\cal H}$ is projected from a testfunction 
$|\Psi_T\rangle$ with a normalization constant ${\cal N}$ and with the 
projection parameter $\theta$. Details of the method are described in 
\cite{HUS97}.

To provide evidence for superconductivity we follow the
standard concept of off diagonal long range order (ODLRO) \cite{YAN62}. 
Therefore we study the vertex correlation function
\[
\label{EqVertexdxmy}
C^V_{d}(r) = 
\frac{1}{L}
\sum\limits_i \sum\limits_{\delta,\delta'} g_\delta g_{\delta'}
\langle c_{i\uparrow}^\dagger c_{i + \delta \downarrow}^\dagger 
c_{i+r + \delta' \downarrow}^{} c_{i+r\uparrow}^{} 
\rangle +
\frac{1}{L} \sum\limits_{i}
\sum\limits_{\delta,\delta'} g_\delta
g_{\delta'}
C^{one}_\uparrow (i,r) C^{one}_\downarrow (i+\delta,r+\delta') 
\]
with the phase factors $g_\delta$, $g_\delta' \pm 1$ to model the 
d-wave symmetry. 
As shown in \cite{MOR94} and in further detail in \cite{HUS94} the d-wave
correlations are positive for larger distances $|r|$ and level off to
a "plateau". Other superconducting symmetries (in particular s-wave) 
fluctuate around zero.
This results has recently been supported by \cite{ZHA97}.
Our current simulations reach the same conclusion for the pure Hubbard
model ($t'=0$). 

The question of superconductivity can be only answered by finite size
scaling.
In the case of weak or intermediate interaction \cite{BOR92}
the behavior of the correlation function is dominated 
by the shell structure of the system. 
Considering the average vertex correlation function 
\[
\label{eqvertex_bar}
\bar{C}_d^{v} \equiv 
\frac{1}{L} \sum_{r} C^{v}_d (r)
\]
with the number of lattice points $L=L_x^2$
standard $1/L_x$ scaling for instance seems to provide clear evidence 
against superconductivity \cite{WHI89}, \cite{IMA91}, \cite{ZHA97}.
In this paper we argue that this conclusion is too simplified.

In this context we introduce a BCS--reduced Hubbard model, the J--model,
with the same 
kinetic Hamiltonian as the tt'--Hubbard model and an interaction
favoring cooper pairs with d-wave symmetry. In momentum space
the Hamiltonian of this model is given by the Hamiltonian
\[
{\cal H}^{BCS} =  
\sum_{k} \sum_{\sigma = \pm 1} \varepsilon_k c_k^\dagger c_k
+ \frac{J}{L} \sum_{{k,p \atop k \ne p}} f_k f_p c_{k,\uparrow}^\dagger
c_{-k,\downarrow}^\dagger c_{-p,\downarrow} c_{p,\uparrow}
\]
with the single particle energies $\varepsilon_k \equiv -2 t \big( \cos(k_x) + 
\cos(k_y )\big) -4 t' \big( \cos(k_x) \cdot \cos(k_y) \big) $
and the form factor $ f_k \equiv \cos(k_x) - \cos(k_y)$ for modeling the
d-wave interaction and $f_k=1$ for the s-wave interaction.
Superconductivity has been rigorously proven for this type of 
BCS-reduced Hubbard models \cite{BUR93}.

\begin{figure}[hbtp]
\begin{center}
\begin{minipage}{12cm}

\epsfxsize 12cm \epsfbox{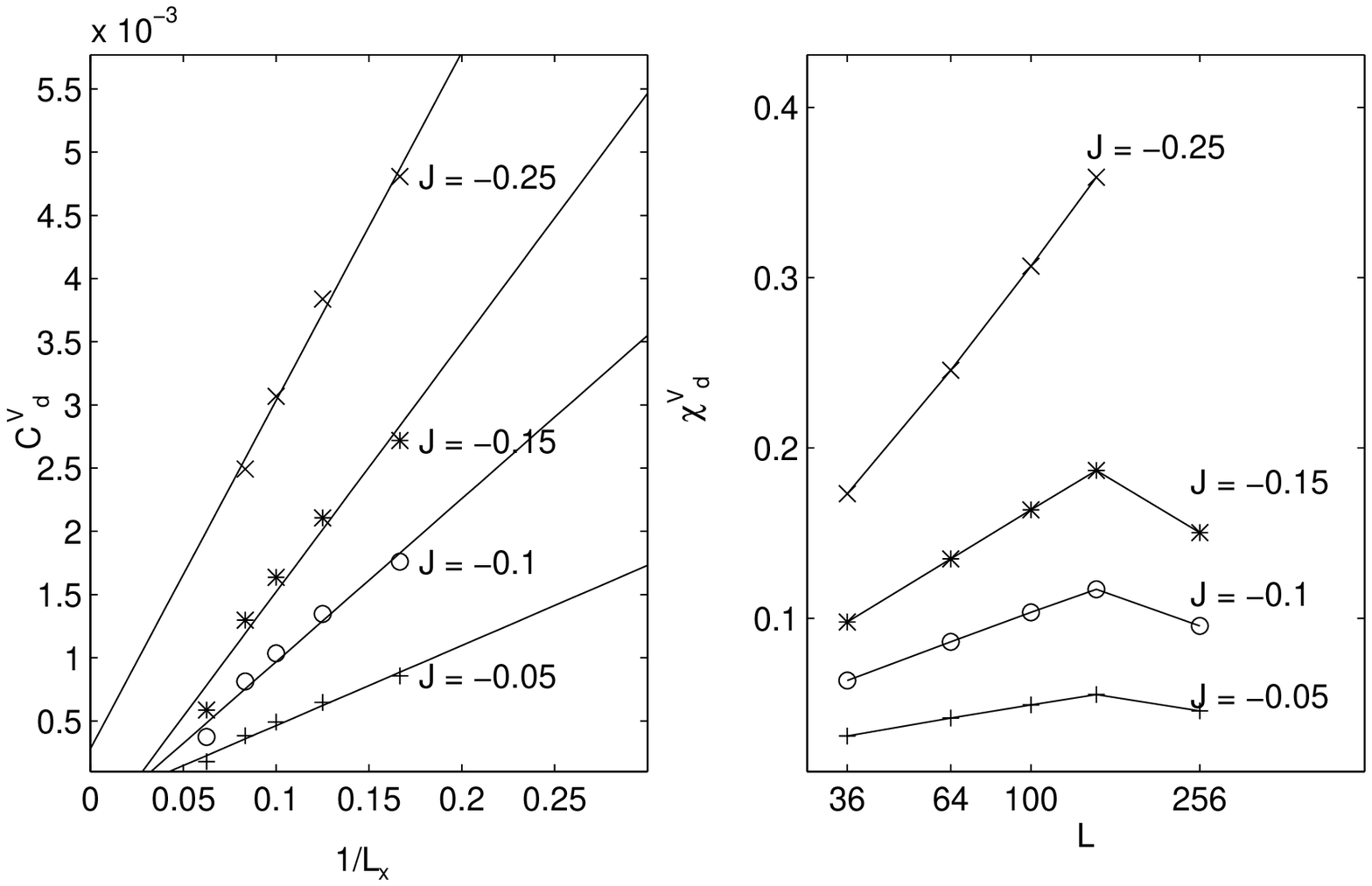}
\end{minipage}
\end{center}
\caption{\label{figscalingtp}
Finite size scaling of the averaged (left) and the cumulated (right) vertex
correlation function. The ground state of the BCS-reduced Hubbard model
with d-wave interaction was calculated with the stochastic 
diagonalization. The filling of table 2 and $t'=0$ was used.
}
\end{figure}

Figure~\ref{figscalingtp} shows the $1/L_x$ scaling of for the J--model.
For weaker interaction we would reach the same conclusion, the absence of 
superconductivity, as other authors in the case of the Hubbard model.
Only the case $J=-0.25$ would be superconducting.
Considering the susceptibility ($\chi^v_d\equiv L \bar{C}^v_d$) again 
for weaker interaction the divergence of $\chi^v_d$ is ambiguous.
This results of figure~\ref{figscalingtp} have been obtained 
with the stochastic diagonalization (SD) \cite{RAE92}, \cite{RAE92/2}, 
\cite{FET97/2}. Details are published in \cite{RAE92}, \cite{FET97/2}.

To avoid these problems with the standard finite size scaling 
ansatz we propose a novel approach \cite{HUS94}. 
Finite size scaling is provided by
comparation of the Hubbard model to the J--model.
Before we carry out this comparation we have to circumvent a further 
complication in the Hubbard model.
The values of the correlations functions $C^v_d(r)$ are extremely large for
smaller distances $|r|$ \cite{MOR94}, \cite{HUS94} and therefore 
susceptible to the fluctuations 
in the numerical simulations. Indeed these fluctuations for smaller distances
exceed the "plateau" value of the $C^v_d(r)$.
As only the long range behavior is of interest for superconductivity
we restrict the average vertex correlation function 
\[
\label{eqvertex_p}
\bar{C}^{v,p}_s \equiv 
\frac{1}{L_c} \sum\limits_{{r \atop |r|>|r_c|}} C^v_s(r)
\]
to the distances $|r| > |r_c|$. In equation~\ref{eqvertex_p} $|r_c|$ is
a critical distance and $L_c$ is the number of points with
$|r| > |r_c|$ for $r \in {1,\ldots L}$. 
Typically we choose $|r_c| = 1.9$.

We now make the assumption, that 
the same finite size scaling behavior of the superconducting 
correlation functions $\bar{C}^{v,p}_{d}$ is equal for both models. 
Accordingly we have for each interaction $U$ and system size $L$ an
effective $J_{eff}$ of the J--model.

We determine $J_{eff}$ in the following way:
For the system parameters $L$, $\langle n\rangle$ (filling) 
and $t'$ and the 
interaction $U$ we calculate $\bar{C}^{v,p}_d$ for
the Hubbard model with PQMC.
For the same set of parameters 
we tune $J$ using the SD method to obtain the same 
value $\bar{C}^{v,p}_d$ in the J--model.
This $J$ is our $J_{eff}$.

A first test was carried out for the negative (attractive) Hubbard model,
which is commonly believed to be superconducting (s-wave symmetry).
Results in table \ref{tabeffwwnegu} show a unique $J_{eff}$ for system
sizes 
$6\times 6$ to $12\times 12$ and small deviations at $4\times 4$.
$|r_c| = 1.9$ was chosen for this and all following cases.
It should be mentioned, that the choice of $|r_c|$ is not critical 
for the qualitative behavior of $J_{eff}$.

In table \ref{tabeffwwposu} we return to the repulsive Hubbard model.
For $U=2$ and $t'=-0.22$ we again obtain as in the attractive case a 
unique $J_{eff}$ for system sizes $6\times 6$ to $12\times 12$.
We notice a decrease for $16\times 16$. The same 
effect occurs for the case $U=1$. But for $t'=0$ (the pure Hubbard
model, $U=2$) we find agreement up to $16\times 16$ (table \ref{tabeffwwposu}).

\begin{figure}[hbtp]
\begin{center}
\begin{minipage}{14cm}

\epsfxsize 14cm \epsfbox{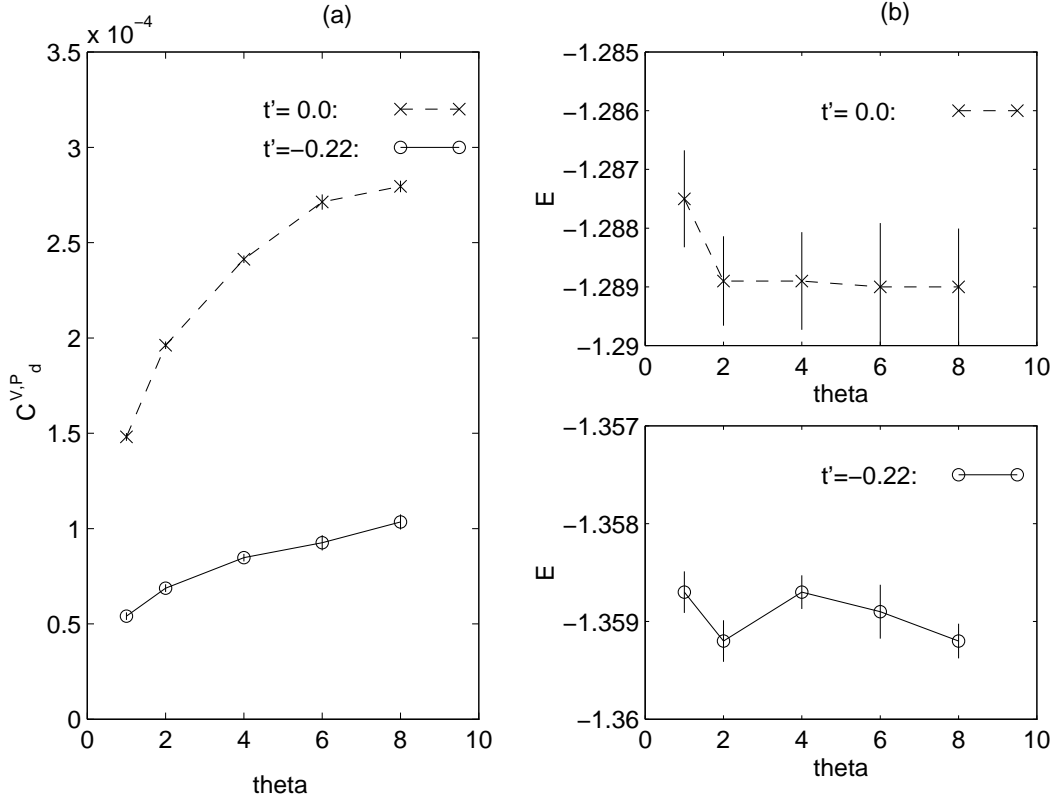}
\end{minipage}
\end{center}
\caption{\label{figtheta_1616}
The $\theta$--scaling is plotted for the $16\times 16$ system
with the filling $n_\uparrow = n_\downarrow = 105$, the interaction 
$U=1$ and $t'=-0.22$ (runs: $t'=-0.22$) and for the $16\times 16$ system
with the filling $n_\uparrow = n_\downarrow = 109$, the interaction 
$U=2$ and $t'=0$ (runs: $t'=0$).
The average sign $\langle {\rm sign}\rangle$ is in both cases 
about one and $\theta /m = \tau =0.125$.
(a) $\theta$ verses ground state energy 
(b) $\theta$ verses the averaged vertex  correlations function with $|r_c|=1.9$ 
$\bar{C}^{V,P}_d$.
}
\end{figure}

This behavior is explained by the existence of different finite size 
gaps. For the small
gaps in the $t'=-0.22$ case the simulations are only valid for relatively
large projection parameters $\theta$, which exceed our numerical 
possibilities. In figure \ref{figtheta_1616} we show the increase
of $\bar{C}^{v,p}_d$ with various $\theta$. In contrast the upper
curve shows the leveling off in the $t'=0$ case for a still moderate
$\theta$. This is caused by the relatively large finite size gap.

Therefore we conclude, that the deviation for $16\times 16$ in the 
$t'=-0.22$ system
is due to insufficiently large $\theta$ in the simulation. The system 
does not reach the ground state properly. Larger $\theta$ are outside
of the reach of methods.
At this point we would like to mention that energy measurements are
still rather insensitive to $\theta$ compared to the vertex correlations \cite{HUS97}.
This is a rather important point as agreement in energy measurements 
was often
used as evidence for the validity of a certain numerical method.
 
Considering again table \ref{tabeffwwposu} we conclude that our
simulations show clear evidence for the existence of d--wave
superconductivity in the Hubbard model.

\begin{figure}[hbtp]
\begin{center}
\begin{minipage}{8cm}

\epsfxsize 8cm \epsfbox{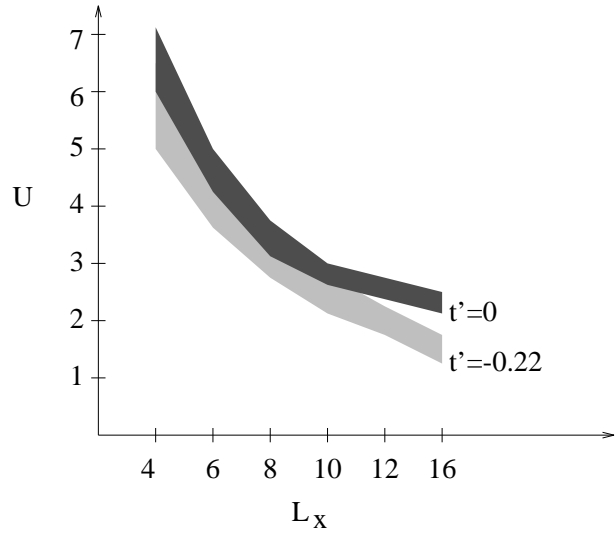}
\end{minipage}
\end{center}
\caption{\label{figgrenzenpqmc}
Limits of the PQMC for the tt'--Hubbard model with the filling $\langle n\rangle \approx 0.8$ and $t'=0$\,/\, $t'=-0.22$
}
\end{figure}

The simulations had to be restricted to values $U\le 2$ and system sizes
up to $16\times 16$ 
because of the convergence problems of the PQMC method
outside this parameter regime. This is clearly indicated by a
dramatic break down of the average sign (table \ref{tabeffwwposu}).
Figure \ref{figgrenzenpqmc} shows the regime of "safe" simulations 
below the shaded areas.

The effective interaction $J_{eff}$ leads to a superconducting $T_c$
in the BCS--model \cite{WHE93}. 
The BCS--$T_c$ has to be considered as at least a rough estimate and 
it does not include fluctuations in the two-dimensional system.
Simulations by Schneider et al. \cite{SCH97} suggest that for the
range of our interactions the deviation of the BCS--$T_c$ and the
real $T_c$ is rather small. Our simulations clearly suggest a 
considerable difference between $T_c^*$ (forming of pairs) and $T_c$
as described by Schneider et al. \cite{HUS94}, \cite{HUS96}.

For the $J_{eff}\approx 0.07$ in table \ref{tabeffwwposu} 
($U=2$, $t'=-0.22$) we only obtain a very low $T_c\approx 1$\,K.
But in a $6\times 6$ system we are able to calculate $U=4$ with a 
sufficient large $\theta$. For $t'=-0.22$ we find $J_{eff} \approx 0.2$.
This effective interaction leads $T_c\approx 30$\,K.
For $t'=0$ we obtain a $T_c\approx 5$\,K. $J_{eff} \approx 0.2$ agrees 
very well with a recently published $12\times 12$ lattice \cite{ZHA97}.
The difference of $T_c$ is explained by the closeness of the Van Hove
singularity in the $t'=-0.22$ case \cite{HUS94}.
Larger values of $\bar{C}^{v,p}_d$ for $U=6$ and $U=8$ as suggested by $4\times 4$
exact diagonalization results may lead to a dramatic increase of $T_c$.
But we do not want base this decision on $4\times 4$ lattice sizes.
The conclusion of \cite{ZHA97}
that the Hubbard model does not exhibit superconductivity for larger $U$
and larger $L$ is not valid. 
The errorbars of $C_d(r)$ are about ten times larger than the value 
of $\bar{C}^v_d$ predicted by our J--model simulations.

In conclusion we provide clear evidence for the existence of d-wave
superconductivity in the Hubbard model. For $U=4$ we obtain a 
$T_c \approx 30$\,K. Therefore the single band Hubbard model has to be 
considered as a serious candidate for the explanation of high $T_c$
superconductivity.

We want to thank T. Husslein, D.M. Newns, H. De Raedt, T. Schneider, 
J.M. Singer and E. Stoll for the helpful discussions and ideas.
This work was supported by the Deutsche Forschungs Gemeinschaft (DFG).
The Leibnitz Rechenzentrum (Munich) grants us a generous amount 
of CPU time on the IBM SP2 parallel computer.



\newpage

\newpage

\newpage

\begin{table}
\begin{center}
{\renewcommand{\arraystretch}{1.2}
\begin{tabular}{|c|c|c|r|r|r|} \hline
$L$ & $n_\uparrow = n_\downarrow$ & $U$ & $t'$ & $\bar{C}^{v}_{s}$ & $J_{eff}$ \\  \hline
$4\times 4$ & 5 & -0.5 & 0.0 & 0.00196(4) & -0.38 \\ \hline
$6\times 6$ & 13 & -0.5 & 0.0 & 0.00089(1) & -0.31 \\ \hline
$8\times 8$ & 25 & -0.5 & 0.0 & 0.00056(2) & -0.30 \\ \hline
$10\times 10$ & 41 & -0.5 & 0.0 & 0.00038(1) & -0.30 \\ \hline
$12\times 12$ & 61 & -0.5 & 0.0 &  0.00028(1) & -0.30 \\ \hline
\hline
$4\times 4$ & 5 & -1 & 0.0 &  0.0045(1) & -0.76 \\ \hline
$6\times 6$ & 13 & -1 & 0.0 & 0.00218(3) & -0.64 \\ \hline
$8\times 8$ & 25 & -1 & 0.0 & 0.00146(2) & -0.63 \\ \hline
$10\times 10$ & 41 & -1 & 0.0 & 0.00105(1) & -0.64 \\ \hline
$12\times 12$ & 61 & -1 & 0.0 & 0.00079(1) & -0.64 \\ \hline
\hline
\end{tabular}
}
\end{center}
\caption{\label{tabeffwwnegu}
Effective interaction $J_{e}$ of the BCS--reduced Hubbard model with 
s-wave interaction. The PQMC has used $\theta=8$ and $m=64$. 
The statistical errors of the last digit are given in the brackets.
}
\end{table}

\newpage

\begin{table}
\begin{center}
{\renewcommand{\arraystretch}{1.2}
\begin{tabular}{|c|c|r|r|r|r|r|} \hline
$L$ & $n_\uparrow = n_\downarrow$ & $U$ & $t'$ & $\langle {\rm sign}\rangle$ & $\bar{C}^{v,p}_{d}$ & $J_{eff}$  \\  \hline
\hline
$6\times 6$ & 13 & 2 & 0.0 & 1.000 & 0.00134(2) & -0.077 \\  \hline
$8\times 8$ & 25 & 2 &  0.0 & 0.999 & 0.00099(2) & -0.075 \\  \hline
$10\times 10$ & 41 & 2 & 0.0 & 0.995 & 0.00073(2) & -0.073 \\  \hline
$12\times 12$ & 61 & 2 & 0.0 & 0.987 & 0.00056(2) & -0.071 \\  \hline
$16\times 16$ & 109 & 2 & 0.0 & 1.000 & 0.000278(6)  & -0.078 \\ \hline
\hline
$6\times 6$ & 13 &  1 & -0.22 &  1.000 & 0.00053(2) & -0.025 \\  \hline
$8\times 8$ & 25 &  1 & -0.22 & 1.000 & 0.00037(1) & -0.025  \\  \hline
$10\times 10$ & 41 &  1 & -0.22 & 1.000 & 0.000259(4) & -0.023 \\  \hline
$12\times 12$ & 61 &  1 & -0.22 & 1.000 & 0.000193(4) & -0.022 \\  \hline
$16\times 16$ & 105 &  1 & -0.22 & 1.000 & 0.000104(4) & -0.016 \\  \hline
$6\times 6$ & 13 &  2 & -0.22 & 0.999 & 0.00134(2) & -0.061 \\ \hline
$8\times 8$ & 25 &  2 & -0.22 & 0.995 & 0.00107(2) & -0.069 \\ \hline
$10\times 10$ & 41 &  2 & -0.22 & 0.973 & 0.00080(2) & -0.068 \\ \hline
$12\times 12$ & 61 &  2 & -0.22 & 0.876 & 0.00063(1) & -0.067 \\ \hline
$16\times 16$ & 105 &  2 & -0.22 & 0.567 & 0.00023(4) & -0.026 \\ \hline
\end{tabular}
}
\end{center}
\caption{\label{tabeffwwposu}
Effective interaction $J_{e}$ of the BCS--reduced Hubbard model with 
d-wave interaction. The PQMC has used $\theta=8$ and $m=64$. 
The statistical errors of the last digit are given in the brackets.
}
\end{table}

\end{document}